
\magnification=1200
\baselineskip=20pt
\tolerance=100000
\overfullrule=0pt

\def\pt {\partial}
\def\At {\overline {A}} \def\Zt {\overline {Z}} \def\Ft {\overline {F}}
\def\nab {\bigtriangledown}\def\O {\Omega}
\def\noi {\noindent}
\def\ov {\over}
\def\sq {\sqrt}

 \def\({\c c} \def\|{\'\i }

\def\sqr#1#2{{\vcenter{\vbox{\hrule height.#2pt
        \hbox{\vrule width.#2pt height#1pt \kern#1pt
          \vrule width.#2pt}
        \hrule height.#2pt}}}}
\def\square{\mathchoice\sqr68\sqr68\sqr{4.2}6\sqr{2.10}6}

\centerline{\bf QUANTIZATION OF MAXWELL-CHERN-SIMONS-PODOLSKY THEORY}

\vskip 2cm

\centerline{A. de Souza Dutra$^1$ and Marcelo Hott$^{2\dagger}$}
\centerline{$^1$Depto. de F\|sica e Qu\|mica, UNESP - Campus de
Guaratinguet\'a}
\centerline{P.O. Box 205, CEP : 12.500, Guaratinguet\'a, S.P., Brazil}
\centerline{e-mail: DUTRA@GRT$\emptyset\emptyset\emptyset$.UESP.ANSP.BR}
\centerline{$^2$Department of Physics and Astronomy, University of
Rochester}
\centerline{Rochester, NY 14627-0171 USA}
\centerline{e-mail: HOTT@URHEP.PAS.ROCHESTER.EDU}
\vskip 3cm

\noindent {\bf \underbar{Abstract}}

We quantize a generalized electromagnetism in 2 + 1 dimensions which
contains a higher-order derivative term by using Dirac's method.  By
introducing auxiliary fields we transform the original theory in a
lower-order derivative one which can be treated in a usual way.

\vskip 3cm

\noindent $^\dagger$On leave of absence from
  Depto. de F\|sica e Qu\|mica, UNESP -
Campus de Guaratinguet\'a, Guaratinguet\'a, S.P., Brazil

\vfill\eject

\noindent {\bf I. INTRODUCTION}

\medskip

In the last years there has been an increasing interest in 2 + 1 dimensions
gauge field theories [1] in part due to the fact that some of these
theories exhibit features which could be associated to the high $T_c$
superconductors and the fractional quantum Hall effect [2].  In such cases
the most studied theories have been those which includes a topological mass
term to the gauge field (Chern-Simons term) that couples to the matter
current.

Besides, the gauge field theories in lower dimensions are a good laboratory
to study phenomena which are not well understood in four dimensions.

Particularly, the quantization of Chern-Simons theories has received much
attention.  Dirac's quantization method [3] was used not only in the
Abelian massive gauge theory [4] but also in the non-Abelian case and in
the pure Chern Simons theory [5].  The Abelian massive gauge theory was
also quantized in a manifestly covariant way [6].

On the other hand, theories with higher-order derivatives have also attracted
considerable interest.  Those terms were already introduced in the past
century by Ostrogradskii.  In the 1940's, aiming to avoid divergences in
Maxwell's theory, Podolsky suggested the inclusion of higher-order derivative
terms [7]. More recently, quantization aspects of this theory has been
analysed by using Dirac's method [8] as well as the Batalin-Fradkin-Vilkovisky
formalism [9]. This interest rely on the possibility of soften the
ultraviolet divergences, leading to a possible attenuation of the problem
of renormalizability for theories like quantum gravity.

Most of the models with higher derivative terms present undesired
properties such as non-renormalizability and tachyonic massive modes.
Despite this, higher-order derivative terms have been used in gravitation
in order to improve the ultraviolet behavior of the Einstein-Hilbert action
[10].  Even in supersymmetry and string theory, higher-order terms play a
certain role.  In string theory, for instance, a term proportional to the
extrinsic curvature of the world sheet was proposed and it has a great
influence in the phase structure of the theory [11].  The idea was
applied to the relativistic particles as well [12].  In supersymmetry,
higher-order derivative terms are a useful regularization, that preserves
supersymmetry.

In this paper we intend to quantize the Maxwell-Chern-Simons-Podolsky
theory by using Dirac's method.  It was shown that a Chern-Simons term can
be generated in Quantum Electrodynamics in 2 + 1 dimension whenever one
integrates the  fermionic fields out [13-15].  As a matter of fact
several other terms can also be generated, including those with
higher-order derivatives of the gauge field.  This can be seen, for
example, in the derivative expansion method [16], like it was done in
reference [14].  In this way, the study of properties of the present
theory can be helpful to the better understanding of gauge effective
actions in 2 + 1 dimensions.

This paper is organized as follows.  In section II we define the model and
derive the propagator.  In section III we map the theory into another one
which does not present higher-order derivative terms and quantize it; in
section IV we discuss the results and make some remarks.

\bigskip

\noindent {\bf II. MAXWELL-CHERN-SIMONS-PODOLSKY THEORY}

\medskip

We start with the following Lagrangian density

$${\cal L} = - {a \over 4}\ F_{\mu \nu} F^{\mu \nu} + {\theta \over 2} \
\varepsilon^{\mu \rho \nu} A_\mu \partial_\rho A_\nu -
{b^2 \over 2}\ \partial^\mu F_{\mu \nu} \partial_\lambda
F^{\lambda \nu} \quad , \eqno(1)$$

\noi where $a,\ \theta \ {\rm and}\ b$ are free parameters which permit us
taking the appropriate limits.  The sign of the Podolsky term has been
considered in conformity with the original work~[7].

In order to obtain the propagator one has to add the gauge-fixing
Lagrangian

$${\cal L}_{gf} = {1 \over 2 \alpha}\ \left( \partial_\mu A^\mu \right)^2
\eqno(2)$$

\noi to the original one, since (1) is not invertible; where $\alpha$ is the
gauge parameter.  In such case the propagator, in momentum space, is given by

$$G_{\mu \nu} = {1 \over [(a+b^2k^2)^2 k^2 - \theta^2]}\
\bigg\{ \big(a+b^2 k^2 \big) g_{\mu \nu} + $$
\smallskip
$$+ \alpha \big[ \big( a+b^2k^2\big) \big( a + {1 \over \alpha} +
b^2 k^2 \big) k^2 - \theta^2 \big] {k_\mu k_\nu \over k^4} +
i \theta \varepsilon_{\mu \nu \rho} {k^\rho \over k^2} \bigg\} \quad ,
\eqno(3)$$

\noi which agrees with reference [9] when the parameter $\theta$ goes to zero.
It is easy to see that the poles of the propagator are defined by the equation

$$y^3 + a_1 y^2 + a_2 y + a_3 = 0 \qquad ; \qquad (y=k^2) \quad ,
\eqno(4)$$

\noi where

$$a_1 \equiv {2 a \over b^2} \quad ; \quad a_2 \equiv {a_1^{\ 2} \over 4} \quad
; \quad a_3 \equiv - {\theta^2 \over b^4} \quad . \eqno(5)$$

\indent From the above equations we can see that the number of massive poles
depends on the choice of the parameters.

By studying the solutions of above equation one sees that, in order to have
only real roots, it is necessary to impose that the discriminant $D\,\equiv\,~
Q^3\,+\,R^2$, where

$$Q\,\equiv\,-\,{a_1^2\over 36}\;,\;R\,\equiv\,{1\over 54}\left({a_1^3\over
4}\,-\,27a_3\right),\eqno(6)$$

\noi be lesser than or equal to zero. So, one have two possibilities:

\noindent {\it i})\quad \underbar {$D\,=\,0$}: Here we have three real
roots, where two are equal (leading to the appearing of ghosts). So we
have that

$$\eqalignno{m_1^2\,&=\,2R^{1\over 3}\,-\,{a_1\over 3},\cr
m_2^2\,&=\,m_3^2\,=\,-\,R^{1\over 3}\,-\,{a_1\over 3}.&(7)\cr}$$

However, the imposition $D\,=\,0$, imply that one must have:
$a_3\,=\,0$, which is a trivial solution with $\theta\,=\,0$; or $a_3\,=\,
{a_1^3\over 54}$, but in this case we get a negative squared mass, revealing
the existence of tachyonic excitations.

\noindent {\it ii})\quad \underbar {$D\,<\,0$}: Now we are faced with three
different real roots of Eq.(4). The solutions can be written as:

$$\eqalignno{m_0^2\,&=\,2\,\rho^{1\over 3}cos\left({\alpha\over
3}\right)\,-\,{a_1\over 3},\cr
m_{\pm}^2\,&=\,-\,\rho^{1\over 3}\left\lbrack cos\left({\alpha\over 3}\right)
\,\pm\,{\sqrt 3}sin\left({\alpha\over 3}\right)\right\rbrack\,-\,{a_1\over
3},&(8)\cr}$$

\noindent where

$$\rho^{1\over 3}\,=\,\left\lbrack R^2\,+\,{1\over 4}\vert a_3^2\,-\,a_3\,
a_1^3/54\vert\right\rbrack^{1\over 2},\eqno(9a)$$

\noindent and

$$\alpha\,=\,arctan\left\lbrack{108{\sqrt {\vert a_3^2\,-\,a_3\,
a_1^3/54\vert}}\over {a_1^3\,-\,108a_3}}\right\rbrack.\eqno(9b)$$

\smallskip

\noindent Besides, the imposition $D\,<\,0$, leaves us with the
possibilities:

\noindent 1) \quad $a_1\,>\,0$:\qquad\qquad\qquad\qquad\qquad\qquad
 $0\,<\,a_3\,<\,{a_1^3\over 54}$

\noindent 2) \quad $a_1\,<\,0$:\qquad\qquad\qquad\qquad\qquad\qquad
${a_1^3\over 54}\,<\,a_3\,<\,0$.

In this case however, it is hard to verify if the above values of $a_3$
and $a_1$ correspond to positive square masses, at least analitically.
Nevertheless, doing a numerical analisys, it appears that the
following rules hold:

\noindent a) For $a_1\,>\,0$, there is no possible way of keeping the
three square masses simultaneously positive.

\noindent b) For $a_1\,<\,0$, one observes that along a reasonable range
of variation for the parameter $a_1$ ($-0.01\,\leftrightarrow\,-100$), that
the region where the three masses are positive, it is always greater than
that where $D\,<\,0$. So one can see that the region defined above
apparently leads to well behaved masses (no tachyons, no ghosts).

Unfortunately, the usual case where $a\,=\,1$, and Podolsky's parameter is
negative, implies that $a_1$ be positive. Consequently the taquyons stays
present in the theory.

\bigskip

\noindent {\bf III.  DIRAC'S QUANTIZATION}

\medskip

In this section we perform the Dirac quantization through two different
ways of gauge fixing the theory.

Looking at the original Lagrangian (1), it is possible to verify that it is
equivalently described by the Lagrangian density

$${\cal L} = -{a \over 4}\ F_{\mu \nu} F^{\mu \nu} + {\theta \over 2}\
\varepsilon^{\mu \rho \nu} A_\mu
\partial_\rho A_\nu + {1 \over 2}\ Z_\mu Z^\mu + {b \over 2}\
F_{\mu \nu} Z^{\mu \nu} \quad , \eqno(10)$$

\noi where $Z_\mu$ is an auxiliary field and $Z_{\mu \nu}$, defined by

$$Z_{\mu \nu} = \partial_\mu Z_\nu - \partial_\nu Z_\mu \quad , \eqno(11)$$

\noi is the associated field strength. For sake of curiosity, it is interesting
to observe that we could also perform a suitable change of variables,
capable of disentangle the term coupling the field strenghts. Thus we would
obtain that

$$\eqalignno{\overline {\cal L}\;&=\;-\,{a\over 4}\Ft_{\mu \nu}\Ft^{\mu \nu}
\;+\;-\,{1\ov 4}\,{b^2\over {\vert a \vert}}\left( 2-sign(-\,a)\right)
\Zt_{\mu \nu}\Zt^{\mu \nu}\;+\;{1\ov 2}\Zt_\mu \Zt^\mu\;+\cr
&+\;{\theta\ov 2}\;\varepsilon^{\mu \nu \rho}\At_\mu\pt_\nu\At_\rho\;-\;
{\theta\ov 4}\,{b^2\ov {\vert a \vert}}\varepsilon^{\mu \nu \rho}\Zt_\mu\pt_\nu
\Zt_\rho\;-\;\theta\,{{\vert b \vert}\ov {\sq a}}\varepsilon^{\mu \nu \rho}
\At_\mu\pt_\nu\Zt_\rho.\cr}$$

In this case it can be seen easily that we would get a Chern-Simons-Maxwell
field interacting with a Chern-Simons-Proca one through crossed
Chern-Simons terms.

However, we will work with the Lagrangian density (10) instead of that
one of equation (1).  Here we do not have to define the momentum associated
with the field time derivative. In any case, the introduction of the auxiliary
fields duplicates the number of variables we are dealing with.

The equations of motion are

$$\eqalignno{a\  \partial_\mu F^{\mu \nu} + \theta\,\varepsilon^{\nu \mu
\alpha}
F_{\mu \alpha} &= b\  \partial_\mu Z^{\mu \nu}\quad , &(12a)\cr
b\  \partial_\mu F^{\mu \nu} &= Z^\nu \quad . &(12b)\cr}$$

Taking the divergence in equation (12b) and using the antisymmetry property
of $F^{\mu \nu}$ we get

$$\partial_\nu Z^\nu = 0 \quad , \eqno(13)$$

which is a Lagrangian constraint.

The equation of motion of the potential $A_\mu$ has the form
$$\left( a - b^2 \square \right) \square A^\nu + 2 \ \theta\,
\varepsilon^{\nu \beta \mu} \partial_\beta A_\mu - \left( a - b^2 \square
 \right)
\partial^\nu \partial^\mu A_\mu = 0 \quad . \eqno(14)$$

\indent From the Lagrangian
 density (10) we have the following primary constraints

$$\eqalignno{\Omega_1 &= \pi_0^{\ A} \approx 0 \quad , &(15a)\cr
\Omega_2 &= \pi^{\ Z}_0 \approx 0 \quad , &(15b)\cr}$$
which are, respectively, the momenta associated to the components $A_0$ and
$Z_0$.

The space-components of the momenta are
$$\eqalignno{\pi^{\ A}_i &= a\ F_{i0} - {\theta \over 2}\ \varepsilon_{ij}
A_j - b\ Z_{i0} \quad , &(16a)\cr
\pi^{\ Z}_i &= -b\ F_{i0}.&(16b)\cr}$$

We can now construct the primary Hamiltonian
$$\eqalign{H_p = &\int d^2 \vec x\Bigl\lbrack\ A_0 \partial_i \pi^{\ A}_i + Z_0
\partial_i \pi^{\ Z}_i - {\pi^{\ A}_i \pi^{\ Z}_i \over b}
- {a \over 2b^2}\ \big( \pi^{\ Z}_i \big)^2 +\cr
\noalign{\vskip 4pt}%
&+ {\theta \over 2b}\ \varepsilon_{ij} \pi^{\ Z}_i A_j - {\theta \over 2}\
\varepsilon_{ij} A_0 \partial_i A_j - {Z^{\ 2}_0 \over 2} + {Z^{\ 2}_j
\over 2} + {a \over 4}\ F_{ij}^{\ 2} - \cr
\noalign{\vskip 4pt}%
&\qquad\quad - {b \over 2} \ Z_{ij} F_{ij} + \lambda_1 \Omega_1 + \lambda_2
\Omega_2\Bigr\rbrack \quad , \cr}\eqno(17)$$
where $\lambda_i$ are the Lagrange multipliers.

Since the primary constraints must be maintained in the time,
their consistency conditions generates two other constraints (secondary
constraints).
$$\Omega_3 = \dot \Omega_1 = \partial_i \pi_i^{\ A} -
{\theta \over 2}\ \varepsilon_{ij} \partial_i A_j
\approx 0 \eqno(18a)$$
and
$$\Omega_4 = \dot \Omega_2 = \partial_i \pi^{\ Z}_i - Z_0 \approx 0 \quad .
\eqno(18b)$$

The definition of $\pi^{\ Z}_i$ in equation (16b) and equation (18b) is
nothing more than one of the Lagrangian constraints of equation (12b).
Furthermore, equation (18b) can be seen as a \lq\lq Gauss law with
sources".
$$\partial_i E_i \approx - {Z_0 \over b} \quad , \eqno(19)$$
where $E_i = F_{i0}$.

It is possible to verify that the consistency of $\Omega_3$ is identically
fulfilled,
$$\dot \Omega_3 \equiv 0 \quad . \eqno(20)$$
On the other hand, consistency of $\Omega_4$ gives a condition to
 $\lambda_2$
$$\dot \Omega_4 = - \partial_i Z_i + \lambda_2 \approx 0$$
$$\lambda_2 \approx \partial_i Z_i \quad . \eqno(21)$$

Up to now we have four constraints.  $\Omega_2$ and $\Omega_4$ are second
class constraints and $\Omega_1$ and $\Omega_3$ are first class.  In order
to transform those last into second class constraints we have to do a
gauge choice. This can be done through two different ways. One can
introduce additional second class constraints, or break the gauge
invariance directly at Lagrangian level by using the Faddeev-Popov's trick.
Here we will apply the two approaches respectively.

Equation (14) suggests the following gauge choices
$$\left( a - b^2 \square \right) \partial_i A_i = 0 \quad , \eqno(22)$$
$$\partial_i A_i = 0 \quad . \eqno(23)$$
We choose the last one, which is going to be our fifth constraint
$$\Omega_5 = \partial_i A_i \approx 0 \quad , \eqno(24)$$
whose consistency generates a sixth constraint
$$\Omega_6 = \dot \Omega_5 = \nabla^2 A_0 +
{\partial_i \pi^{\ Z}_i \over b} \approx 0 \quad , \eqno(25)$$
that, in its turn, gives a condition to $\lambda_1$
$$\dot \Omega_6 = \nabla^2 \lambda_1 - {\partial_i Z_i \over b} \approx 0
\quad , \eqno(26)$$
and no more constraints are generated at all.

It is worth mentioning that the gauge choice (23) can be satisfied if we
choose the gauge function as [5]
$$\Lambda = - {1 \over \nabla^2} \ \partial_i A_i \quad , \eqno(27)$$
for, starting from the gauge transformation
$$A^\prime_j = A_j + \partial_j \Lambda \eqno(28)$$
we get
$$\partial_j A^\prime_j = \partial_j A_j
+ \nabla^2 \Lambda = 0 \quad , \eqno(29)$$
and, since
$$A^\prime_0 = A_0 + \partial_0 \Lambda \quad , \eqno(30)$$
we have
$$\nabla^2 A^\prime_0 = \nabla^2 A_0 - \partial_0 \partial_i A_i =
\pt_i F_{0i} = - {\partial_i \pi^{\ Z}_i \over b} \quad , \eqno(31)$$

\noi where we have used equation (16b).  This way we recover the constraint
$\Omega_6$, when this is strongly imposed,

\smallskip

$$\nabla^2 A^\prime_0 + {\partial_i \pi^{\ Z}_i \over b} = 0
\eqno(32)$$

\smallskip

Now we have only second class constraints and we can invert the constraint
matrix.  The commutation relations among the constraints are
$$\eqalignno{\big\{ \Omega_1 (\vec x), \Omega_6 (\vec y) \big\} &=
-\nabla^2 \delta (\vec x- \vec y) \quad , &(33a)\cr
\big\{ \Omega_2 (\vec x), \Omega_4 (\vec y) \big\} &=
 \delta (\vec x- \vec y) \quad , &(33b)\cr
\big\{ \Omega_3 (\vec x), \Omega_5 (\vec y) \big\} &=
-\nabla^2 \delta (\vec x- \vec y)  &(33c)\cr}$$
and the others vanish.  The inverse of the constraint matrix
$$C(\vec x - \vec y)^{-1}_{ij} = \pmatrix{ && &0 &0 &\nabla^{-2}\cr
\noalign{\vskip 5pt}%
 & 0 &  &-1 &0 &0\cr
\noalign{\vskip 5pt}%
& &  &0 &\nabla^{-2} &0\cr
\noalign{\vskip 5pt}%
0 &1 &0 & & &\cr
\noalign{\vskip 5pt}%
0 &0 &-\nabla^2  & &0 &\cr
\noalign{\vskip 5pt}%
-\nabla^2 &0 &0 &&&\cr}
\delta (\vec x - \vec y) \quad , \eqno(34)$$
and the only Dirac brackets among the dynamical
variables which do not vanish are
$$\eqalignno{&\big\{ A_i (\vec x),\pi^{\ A}_j \big\}_{\rm D.B.} \quad\,=
- \big( \delta_{ij} - {1 \over \nabla^2}\ \partial_i \partial_j \big)
\delta (\vec x - \vec y) \quad , &(35a)\cr
&\big\{ Z_i (\vec x),\pi^{\ Z}_j (\vec y) \big\}_{\rm D.B.} =
-  \delta_{ij} \delta
 (\vec x - \vec y) \quad , &(35b)\cr
&\big\{ Z_i (\vec x), A_0 (\vec y) \big\}_{\rm D.B.} \;=
-  {1 \over b}\
{1 \over \nabla^2}\   \partial_i
\delta (\vec x - \vec y) \quad , &(35c)\cr
&\big\{ Z_i (\vec x), Z_0 (\vec y)  \big\}_{\rm D.B.} \;=
 \partial_i
\delta (\vec x - \vec y)  &(35d)\cr}$$

and

$$\left\{ \pi^{\ A}_i (\vec x) , \pi^{\ A}_j
(\vec y) \right\}_{\rm D.B.} =
{\theta \over 2} \left( \varepsilon_{jk} {1 \over \nabla^2}\
\partial_k \partial_i -
\varepsilon_{ik} {1 \over \nabla^2}\
\partial_k \partial_j \right) \delta (\vec x - \vec y)
\eqno(35e)$$

\indent From equation (16b) and the Hamilton equation for the momentum~
$\pi^{\ Z}_i$,
$$\dot \pi^{\ Z}_i = Z_i + b\ \partial_k F_{ki} \quad , \eqno(36)$$
we get the \lq\lq Ampere law in the presence of a source"
$$\dot E_i + \varepsilon_{ki} \partial_k B = - {Z_i \over b} \quad ,
\eqno(37)$$

\noi where $B = \varepsilon_{ij} \partial_i A_j$ is the magnetic field.

Now we perform the Dirac quantization by following the second approach
quoted before. For this we add the following gauge fixing term in the
Lagrangian density (10),

$${\cal L}_{GF}\;=\;{\alpha\over {2}}\sigma^2\;+\;\sigma\;(\pt_\mu A^\mu),
\eqno(38)$$

\noi where $\sigma$ is an auxiliar field, used to permit that the Lorentz
condition be introduced as a linear constraint. Its introduction produces the
following set of primary second class constraints:

$$\O_1\;=\;\pi_0^A-\sigma\;;\;\O_2\;=\;\pi_0^Z\;;\;\O_3\;=\;\pi_\sigma,
\eqno(39)$$

\noi whose preservation in time generates the secondary constraints:

$$\eqalignno{&\O_4\;=\;\alpha\;\sigma - {\vec \nabla\cdot \vec A}\;;\;
\O_5\;=\;{\vec
\nabla\cdot \vec \pi^Z}\;-\;Z_0;\cr
&\O_6\;=\;\alpha\;{\vec \nabla\cdot \vec A \pi^A}\;+\;{{\alpha\;\theta}\ov 2}\;
\varepsilon^{ij}
\pt_i A_j\;-\;\nabla^2 A_0\;+\;{1\ov {\vert b \vert}}
{\vec \nabla\cdot \vec \pi^Z},
&(40)\cr}$$

\noi and also eliminate the Lagrange multipliers,

$$\nabla^2\lambda_1\;=\;-\,{1\ov {\vert b \vert}}{\vec \nabla\cdot \vec Z}\;;\;
\lambda_2\;=\;
{\vec \nabla\cdot \vec Z}\;;\;\lambda_3\;=\;{\vec \nabla\cdot \vec \pi^A} +
{\theta\ov 2}\;\varepsilon^{ij} \pt_i A_j.\eqno(41)$$

So we finish with six second class constraints, whose inverse matrix is
given by

$$C(\vec x - \vec y)_{ij}^{-1}\;=\;{1\ov 2}\pmatrix {0&0&1&0&0&-\nabla^{-2}\cr
0&0&0&0&-2&0\cr
-1&0&0&\alpha^{-1}&0&0\cr
0&0&-\alpha^{-1}&0&0&-\alpha^{-1}\nabla^{-2}\cr
0&2&0&0&0&0\cr
\nabla^{-2}&0&0&\alpha^{-1}\nabla^{-2}&0&0\cr}\delta({\vec x - \vec y}).
\eqno(42)$$

\noi After using the constraints strongly, we eliminate the fields $A_0\;,\;
\pi_0\;,\;\sigma\;,\;\pi_\sigma\;,\;B_0\;,\;\pi_0^B.$ Obtaining for the
remaining fields the following nonvanishing Dirac brackets:

$$\eqalignno{\left\{ A_i\;,\;\pi_j^A\right\}_{\rm D.B.}\;&=\;-\,\Bigl\lbrack
\delta_{ij}\;-\;
{\pt_i\pt_j\ov \nabla^2}\Bigr\rbrack\;\delta({\vec x - \vec y});&(43a)\cr
\left\{ Z_i\;,\;\pi_j^A\right\}_{\rm D.B.}\;&=\;{1\ov {2\alpha \vert b \vert}}
{\pt_i\pt_j\ov {\nabla^2}}\;\delta({\vec x - \vec y});&(43b)\cr
\left\{ Z_i\;,\;\pi_j^Z\right\}_{\rm D.B.}\;&=\;-\,\delta_{ij}\;\delta(
{\vec x - \vec y});&(43c)\cr
\left\{ \pi^{\ A}_i (\vec x) , \pi^{\ A}_j
(\vec y) \right\}_{\rm D.B.} &=
{\theta \over 2} \left( \varepsilon_{jk} {1 \over \nabla^2}\
\partial_k \partial_i -
\varepsilon_{ik} {1 \over \nabla^2}\
\partial_k \partial_j \right) \delta (\vec x - \vec y).&(43d)\cr}$$

The reduced Hamiltonian is then written as

$$\eqalignno{H_r\;&=\;{1\ov {2\alpha}}({\vec \nab\cdot \vec A})^2\;+\;
\alpha\;({\vec \pi_A})^2\;+\;\alpha\;\theta\;\varepsilon^{ij}\pi_i^A
 A_j\;+\;{1\ov 2}({\vec \nab\cdot \pi^B})^2\;+\cr
&-\;{{\vec Z}^2\ov 2}\;-\;{a\ov {2b^2}}({\vec \pi^Z})^2\;+\;{a\ov 2}\pt_i
A_jF^{ij}\;+\;{\alpha\;\theta\ov {2\vert b \vert}}({\vec A})^2\;+\cr
&+\;{\theta\ov {\vert b \vert}}\varepsilon^{ij}\pi_i^ZA_j\;+\;\vert b \vert
\pt_iZ_jF^{ij}.&(44)\cr}$$

It is be interesting to note that in this case, as might be expected,
the Dirac brackets have some dependence in the gauge parameter.
Furthermore some of the brackets are equal to that obtained using the
former approach to fixing the gauge.

\bigskip

\noindent {\bf IV. CONCLUSIONS}

\medskip

We have performed Dirac's quantization of a generalized electromagnetism in
2 + 1 dimension which contains Maxwell, Chern-Simons and Podolsky's terms.
Our results are in complete agreement with the literature when the proper
limits of the parameters are taken.  Moreover the freedom to choose these
parameters allow us to have distinct massive poles for the fields,
including the existence of tachyonic modes which, unfortunately, can not be
eliminated consistently.

We used the auxiliary field method to reduce the model to a lower-order
derivative one and consequently pay the price of duplicating the number of
dynamical variables.  This is an equivalent price one would have to pay if
auxiliary field were not used, since one has to consider the
time-derivative field as well as its conjugated momentum as independent
variables.

As a consequence of introducing auxiliary fields we can see the Maxwell
equations, equations (19) and (37), in a different point of view.  The
higher-derivative fields can be seen as sources to the fields themselves.

If one considers the effective action to the gauge field generated from
the integration of
fermion fields, the theory treated here should present interesting
features.  Since Chern-Simons and Podolsky's terms can be generated
dynamically it would be of interest to analyse the dependence with the
temperature of Chern-Simons [14,15] and Podolsky's parameters, consequently
the spectrum of massive excitation.

\bigskip

\noindent {\bf ACKNOWLEDGEMENTS}

\medskip

One of the authors (M.H.) would like to thank Prof. A. Das for suggestions
and the hospitality and support during his stay at the Department of Physics
and Astronomy, University of Rochester, NY.  This work was also supported
by Funda\c c\~ao de Amparo a Pesquisa no Estado de S\~ao Paulo (FAPESP)
 and Conselho Nacional de Desenvolvimento Cient\|fico e Tecnol\'ogico
 (CNPq).

\vfill\eject

\noindent {\bf REFERENCES}

\medskip

\item{1.} S. Deser, R. Jackiw and S. Templeton: Ann. Phys. (N.Y.) {\bf 140}
(1982) 372.
\item{  } C. R. Hagen: Ann. Phys. (N.Y.) {\bf 157} (1984) 342.

\item{2.} F. Wilczek: Fractional Statistics and Anyon Superconductivity,
Singapore: World Scientific, 1990.

\item{  } S. Forte: Rev. Mod. Phys. {\bf 64} (1992) 193.
\item{3.} P. A. M. Dirac: Lectures on quantum mechanics, New York: Yeshiva
University, 1964.

\item{  } K. Sundermeyer: Constrained dynamics, Lecture Notes in Physics,
vol. 169.  Berlin, Heidelberg, New York: Springer Verlag, 1982.

\item{4.} J. M. Martinez-Fern\'andez and C. Wotzasek: Z. Phys. C {\bf 43}
(1989) 305.

\item{5.} Qiong-gui Lin and Guang-jiong Ni: Class. Quantum Grav. {\bf 7}
(1990) 1261.

\item{6.} T. Kimura: Prog. Theor. Phys. {\bf 81} (1989) 1109.

\item{7.} B. Podolsky: Phys. Rev. {\bf 62} (1942) 68.

\item{8.} C. A. P. Galv\~ao and B. M. Pimentel: Can. J. Phys. {\bf 66}
 (1988) 460.

\item{9.} J. Barcelos-Neto, C. A. P.
Galv\~ao and C. P. Natividade: Z. Phys. C {\bf 52} (1991) 559.

\item{10.} K. S. Stelle: Phys. Rev. D{\bf 16} (1977) 953.

\item{11.} A. Polyakov: Nucl. Phys. B {\bf 268} (1986) 406.

\item{12.} R. D. Pisarki: Phys. Rev. D{\bf34} (1986) 670.

\item{13.} A. Redlich: Phys. Rev. D{\bf29} (1984) 2366.

\item{14.} K. S. Babu, Ashok Das and Prasanta Panigrahi: Phys. Rev.
D{\bf36} (1987) 3725.

\item{15.} I. J. R. Aitchison, C. D. Fosco and C. Taylor:
Phys. Rev. D{\bf 48} (1993) 5895.

\item{16.} C. M. Fraser: Z. Phys. C {\bf 28} (1985) 101.

\end